\documentclass[sigplan,screen, nonacm]{acmart}

\AtBeginDocument{%
  }

\copyrightyear{2022}
\acmYear{2022}
\acmDOI{XXXXXXX.XXXXXXX}

\acmConference[JDCDGGG '22]{}{September 9--11,
  2022}{Virtual, Japan}

\usepackage{subfiles}
\usepackage[utf8]{inputenc}
\usepackage{epsdice}
\usepackage{amsfonts}
\newcommand\vcdice[1]{\vcenter{\hbox{\epsdice{#1}}}}
\usepackage{xcolor}
\usepackage{pgfplots}
\usepackage{pgfplotstable}
\pgfplotsset{compat=1.17}

\makeatletter
\pgfplotsset{
    /pgfplots/flexible xticklabels from table/.code n args={3}{%
        \pgfplotstableread[#3]{#1}\coordinate@table
        \pgfplotstablegetcolumn{#2}\of{\coordinate@table}\to\pgfplots@xticklabels
        \let\pgfplots@xticklabel=\pgfplots@user@ticklabel@list@x
    }
}
\makeatother

\usetikzlibrary{positioning,fit,calc}
\tikzset{block/.style={draw,thick,text width=2cm,minimum height=1cm,align=center},
         line/.style={-latex}
}
\usepackage{svg}
\usepackage{listings}

\usepackage{mdframed}

\begin{document}

\title{GNOLL: Efficient Software for Real-World Dice Notation and Extensions }

\author{Ian Frederick Vigogne Goodbody Hunter}
\email{ianfhunter@gmail.com}
\orcid{0000-0003-3408-8138}
\affiliation{%
  \institution{Independent Researcher}
  \city{Kells}
  \state{Meath}
  \country{Ireland}
}
\renewcommand{\shortauthors}{Hunter, IFVG}

\begin{abstract}
    GNOLL (\lq GNOLL's Not *OLL\rq ) is a software library for dice notation --- A method of describing how to roll collections of dice. Unlike previous papers, GNOLL's dice notation syntax is focused on parsing a language that tabletop role-players and board gamers already use for specifying dice rolls in many popular software applications. Existing implementations of such a syntax are either incomplete, fragile, or proprietary, meaning that anyone hoping to use such syntax in their application likely needs to write their own solution. GNOLL is an open-source project using the compilation tool \lq YACC\rq\  and lexical tool \lq LEX\rq\  which can be integrated into many applications with relative ease. This paper explores GNOLL's extended dice notation syntax and its competitive performance.
\end{abstract}


\begin{teaserfigure}
  \includegraphics[width=\textwidth]{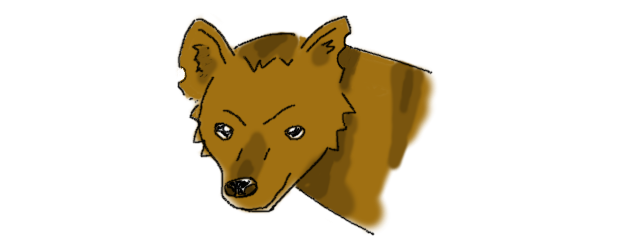}
  \caption{The author's rendition of a hyena/gnoll.}
  \Description{The author has drawn a picture of a hyena/gnoll.}
  \label{fig:teaser}
\end{teaserfigure}

\maketitle


\section{Introduction}

\subsection{Paper Overview}
This paper describes GNOLL, a software library for interpreting a DSL (Domain Specific Language) for dice notation. 

GNOLL is a software library for interpreting a DSL (Domain Specific Language) for an extended set of dice notation. This paper explores GNOLL's supported operations, its performance and provides a much-needed implementation reference for other dice notation interpreters. GNOLL is licensed under GNU General Public License v3.0.

\subsubsection{What's in a name?} 

There are not many publications specifically tackling this issue and the ones that are (as discussed later in section \ref{background_work}) are named  ROLL \cite{ROLL} and TROLL \cite{TROLL} --- To distinguish this research and still pay homage to the original work, GNOLL is a recursive acronym expanded from \lq GNOLL's Not *OLL\rq .

In fantasy-based tabletop role-playing games (TTRPGs) settings, a player might come across a gnoll in their adventure --- they are a hyena-human hybrid typically on the side of evil.


\subsubsection{Paper Layout} 

In this introductory section, we address a few basic questions such as \lq Why is such a project needed?\rq\  and \lq Aren't there existing solutions that can be used?\rq. The architecture of GNOLL and some of the parsing decisions are then discussed in Section \ref{technical_details}.

A collection of results are presented in the subsequent section and the paper is concluded with a summary of GNOLL's impact and a discussion on its merits, limitations, and desirable future work for the project. A list of references and an appendix follow.

\subsection{Demand for a solution}

Though a popular pastime in the 1970s and 1980s, TTRPGs experienced a fall from grace after a moral panic over the percieved content of the games (colloquially known as the "Satanic Panic") \cite{SatanicPanic,Waldron2005RolePlayingGA}. However, over the last decade, there has been a resurgence in their popularity \cite{RPGgrowth}. The popular \lq Dungeons \& Dragons\rq\ TTRPG was recently reported by Hasbro to have had its ninth year of consecutive growth \cite{hasbro2021}. 

In this new era of tabletop gaming, many aspects of the hobby have changed --- revisions of system rules, increased inclusiveness \cite{diversity_and_dragons} and the proliferation of software tools to assist \lq Dungeon Masters\rq\ and their players. During the international COVID-19 pandemic, many groups shifted to play wholly online as they could no longer meet in person. Opinions gathered on the online platform Reddit suggested that while many missed the physical presence of their adventuring companions, they appreciated the tools available for online play --- though a large cohort complained about their instability and complexity of use \cite{covid19DND} .

Many of these tools use common dice notation, but standards differ between platforms and the software used for parsing this notation is predominantly prohibitive to collaboration --- The majority of parsers are closed-source or have their code beneath a server-side application programming interface (API). Available open-source solutions have other problems, many of them supporting only a limited set of dice notation extensions, too tightly coupled with a related product (e.g. character sheets) or left unmaintained.

Because of this, if someone wished to create a new software tool for the hobby they would likely need to create their own dice notation parser. GNOLL provides a solution to this problem by offering a open source software that can be easily integrated into a new (or existing) product.

\subsection{Previous Work}
\label{background_work}

In TROLL\cite{TROLL} (The successive publication to ROLL\cite{ROLL}), Mogensen presents a machine-friendly DSL for rolling dice (Figure \ref{fig:trollcode}). It is a reasonably approachable syntax for those familiar with other programming languages, containing many of the same structures --- loops, conditionals, assignments, etc. However, these constructs are less accessible to users unfamiliar with programming languages and because of this there would likely be a separation between novice and advanced users should a TTRPG tool wish to integrate this. 

TROLL's usage has been primarily used by researchers interested in probability calculation, rather than direct usage in TTRPG software\cite{TROLL}. Dungeons \& Dragons released statistics in 2019 revealing that 12\% of players were age 8 to 12, furthermore 40\% of players were age 24 or younger \cite{DND_2019}. It is clear that because there are many players who are not programmers and even who are not in high school yet, that there is a need for notation to be straightforward and intuitive.


\begin{figure}[h]
    \centering{}
    \begin{mdframed}
\begin{lstlisting}[columns=fullflexible]
largest M N#(sum accumulate x := d10 until x < 10)
\end{lstlisting}
    \end{mdframed}
    \caption{An example of TROLL's Syntax}
    \label{fig:trollcode}
\end{figure}


As was the case in 2009 when TROLL was published, standards on dice notation or publication thereof continue to be scarce. There are documentation listings of particular implementations\cite{avrae_dice,foundryVTTdice,roll20_dice}, but a distinct lack of resources which explain those notation decisions, suggest standardization or encourage co-development.

Several dice notation parsing libraries are available on different programming languages' package managers, however many of these are biased to the more popular TTRPG systems or are short-term effort projects without much consideration given to long term support (e.g. Several of these use Regular Expressions (RegEx) to parse their syntax which can be prone to error and ill performance \cite{DBLP:journals/corr/abs-1210-4992}). The most readily usable and complete implementation the author has been able to find has been RPG Dice Roller \cite{RDR} (There tends to be a lack of distinctive or unique names in this space!), but while it supports a wide range of dice notation syntax, it does not support symbolic dice and only targets a single language (JavaScript) which limits integration options.

Publicly available implementations of dice notation interpreters generally are focused on a single end use-case and to be usable by different programming languages would require significant extra effort by a developer.

\section{Technical Details}
\label{technical_details}

\subsection{Code Architecture}
\label{code_architecture}

GNOLL was written in the C programming language to allow for easier bindings to other languages through software like SWIG\cite{swig_paper} or CPPYY \cite{cppyy_paper}. C has the added advantage of enabling performant code, which allows users to issue more complex dice rolls in an allowed execution time.

Specifically, GNOLL is built upon LEX \cite{LEX} --- a lexical analyser generator, and YACC \cite{YACC} --- a LALR ("Look Ahead Left-to-Right") Context-Free Grammar (CFG) parser generator. The main contribution of GNOLL is contained in this CFG and the functions called as it is parsed.

\begin{figure}
    \centering
    \begin{tikzpicture}
      \node[block, yshift=-4cm] (a) {Statement Resolution};
      \node[draw] at (3,-4) {"(15)"};

      \node[block, yshift=-3cm] (b) {Programmatic Operations};
      \node[draw] at (3,-3) {"(15)"};

      \node[block, yshift=-2cm] (c) {Mathematical Operations};
      \node[draw] at (3,-2) {"[(13)+(2)]"};

      \node[block, yshift=-1cm] (d) {Dice Operations};
      \node[draw] at (3,-1) {"[(4,13)kh] + (2)"};

      \node[block, yshift=-0cm] (e) {Dice Rolls / Terminals};
      \draw[line](-2,0) -- (-2,-4);
      \node[draw] at (3,0) {"[2d20]kh+[2]"};
    \end{tikzpicture}

    \caption{ A simplified deconstruction of how the roll "2d20kh+2" is interpreted by GNOLL}
   \label{fig:hierarchy}
\end{figure}
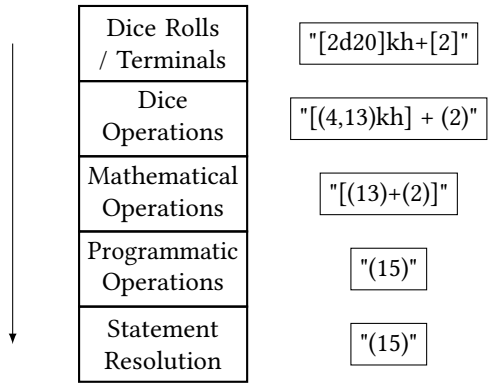

A high-level diagram of how the GNOLL interpreter breaks down a statement can be seen in Figure \ref{fig:hierarchy}.
Characters in the input string are processed into tokens by LEX before they are parsed through the grammar. The underlying dice rolls and values are resolved first, before applying the \lq dice operations\rq\ (such as rerolling, removing dice from a pool, etc. A complete list of GNOLL's supported dice operations are shown in Section \ref{diceOps}). Simple mathematical operations are then resolved, before higher level operations and the final result output.

For more information, there are a detailed set of railroad diagrams depicting the grammar in Appendix \ref{ax:grammar}. 

Whilst GNOLL has made an effort to choose the most appropriate token symbols, there have been decisions made for some rarer symbols to not be supported. Thankfully, if one wishes to fork the GNOLL project, LEX makes it simple to change the tokens associated with each property.

Currently, GNOLL has bindings to Python (via CPPYY\cite{cppyy_paper}) and Perl (via SWIG). Releases of the Python interface are published to the Python Package Index (PyPI) \cite{pypi}. The core of the library is written in C and can be imported to C++ projects. It is the ambition of the project to further expand the language support in the future.

\subsection{Scope of GNOLL}
\label{scope}
GNOLL is intended to be a comprehensive interpreter of dice notation and extensions thereof. It is \textit{not} expected to integrate deeply with TTRPG character sheets, GUI display formatting, or other secondary processing that dependent software may wish to perform.

GNOLL is intended to be primarily a library for inclusion in other projects, though it may contain its own sample applications, test interfaces and scripts.

\subsection{Notation Interpretation}
\label{notation}

GNOLL's syntax was informed by the various documentation of existing dice notation platforms, by Mogensen's paper on RPG Dice Mechanisms\cite{dice_rolling_mechanisms}, and through the notation used in various TTRPG sourcebooks.

\subsubsection{Dice Types}

Internally, GNOLL understands two types of dice --- numeric and symbolic.

\paragraph{Numeric Dice}
\label{simple_dice}
Throughout the different literature, there is consistency about the most basic dice roll form:

\[  x\textbf{d}y,\ where\ x,y\ \in{} \mathbb{Z}^{+}  \]

x represents the amount of dice to roll and y, the number of sides of these dice. An absence of the x term implies a value of 1. Negative values of x will be treated as a negation of the overall roll. Omitting the y term will often emit an error, but some systems will default to a six-sided dice (e.g. the Scum \& Villiany RPG \cite{SaV})- the most common dice in most board games and tabletops. Negative values of y produce an error as dice with a negative number of sides are physically impossible. Some examples of these statements are shown in Equations \ref{eq:dice},\ref{eq:neg},\ref{eq:err}.

\begin{equation}
  d6 = \{ \vcdice{3} \} = 3 
   \label{eq:dice}
\end{equation}
\begin{equation}
  -1d6= -\{\vcdice{5}\}  = -5 
   \label{eq:neg}
\end{equation}
\begin{equation}
d-6=ERROR
   \label{eq:err}
\end{equation}

Where multiple dice exist, these values are generally added together to create a larger value. In Equation \ref{eq:twodice} two dice are rolled and summed. If one wished to keep the rolls distinct in the final result this can be distinguished by a separator token (in GNOLL, a semi-colon) as seen in Equation \ref{eq:seperator}.

\begin{equation}
  2d6=\{ \vcdice{2}, \vcdice{6} \} = 8
  \label{eq:twodice}
\end{equation}
\begin{equation}
d6;d6=\{\{ \vcdice{2}\},\{ \vcdice{6} \}\} = 2,6
\label{eq:seperator}
\end{equation}

GNOLL tokens are all lowercase. This choice was made to reduce conflicts with macros (Section \ref{macros}) which are all upper-case. One example of the need for this rule is the misleading \textbf{D66} notation used in Game Workshop's Mordheim\cite{d66} which rolls two d6 dice and uses them as if they were concatenated, rather than being a 66-sided dice as one might expect. The difference can be seen in Equations \ref{eq:d66_1} and \ref{eq:d66_2} --- it is impossible for a Mordheim-style dice result to be a single digit.

\begin{equation}
\textbf{D66}=\{ \vcdice{3}\times{}10+ \vcdice{1}\}  = 31
\label{eq:d66_1}
\end{equation}
\begin{equation}
d66=\{ \vcdice{5}\} = 5
\label{eq:d66_2}
\end{equation}

To allow more flexibility, GNOLL allows a user to specify any combination of faces through an extended syntax that allows individual number selection or a numeric range. Equation \ref{eq:arb_dice} shows a 11 sided dice where the values range from 1 to 10 and an additional 100 value.

\begin{equation}
d\{1,2,3..8,9,10,100\}=\{ \vcdice{3} \}  = 3
\label{eq:arb_dice}
\end{equation}

\paragraph{Symbolic Dice}

GNOLL also supports symbolic dice --- dice that do not have a strictly numeric value associated with them. A Fate Die --- a six-sided die with two blank sides, two sides with a plus symbol, and two sides with a minus symbol -- can be represented as shown in Equation \ref{eq:fate}, but this is also commonly referred to with the shorthand "df" and this representation is supported also. The symbolic syntax can support strings of up to 100 characters. To make it easier to use, these can be stored as macros (Section \ref{macros}) for less verbose statements.

\begin{equation}
    \label{eq:fate}
    d\{ \text{`-', `-', `0', `0', `+', `+'}\} = \text{`+'}
\end{equation}

\paragraph{Special Dice}
\label{special_dice}

GNOLL supports some common shorthands for different dice types.
\begin{itemize}
    \item $ d\% \rightarrow{} d100 $
    \item $ c \rightarrow{} d\{`HEADS', `TAILS'\} $
    \item $ df \rightarrow{} d\{`-', `-', `0', `0', `+'\} $
\end{itemize}

\subsubsection{Dice Operations}
\label{diceOps}
\paragraph{Drop/Keep Notation}

One of the more common scenarios in a tabletop game is to roll two or more dice and choose the higher or lower of the dice to apply a bonus/penalty to the player (sometimes called \lq Advantage\rq\ and \lq Disadvantage\rq\  respectively). 

The more general operation is to "keep" the highest dice or to "drop" the lowest. Sadly, there is much divergence in syntax across various dice rolling platforms (Table \ref{tab:dropkeep}).




\begin{table}[h]
  \caption{Keep Syntaxes in various dice rollers}
  \label{tab:dropkeep}
  \begin{tabular}{ccl}
    \toprule
    Platform & Keep Highest & Keep Lowest\\
    \midrule
     Roll20 \cite{roll20_dice} & 2d6kh & 2d6kl  \\ 
     Rolz \cite{rolz} & 2d6h & 2d6l \\
     OpenRoleplaying.org \cite{oRP} & 2d6-L & 2d6-H  \\
     RoleGate \cite{rolegate} & 2d6k & 2d6kl  \\
     Dice.Run \cite{dicerun} & 2d6k & 2d6d  \\
     FoundryVTT \cite{foundryVTTdice} & 2d6max & 2d6min  \\
     Avrae \cite{avrae_dice} & 2d6k1 & 2d6p1  \\
  \bottomrule
\end{tabular}
\end{table}

GNOLL supports a syntax as follows to be applied after a dice definition.
\[ \lbrack kd \rbrack \lbrack hl\rbrack\ z\ where\ z\ \in{} \mathbb{Z}^{+} \]
Z represents the amount of values to drop or keep from the original amount. An absence of z implies a value of 1. 

GNOLL uses this syntax primarily to avoid ambiguity, but also because the terms \lq keep\rq\  and \lq drop\rq\  are commonly used in documentation that describes this feature. Where using `d' for dropping dice has been proposed to these systems, there is often a parsing conflict with the core `d' term used for defining a dice (e.g. \textbf{d}6). GNOLL does not have this issue as it requires `l' or `h' to be specified afterwards.

\begin{equation}
2d6kh=\{ \vcdice{3}, \vcdice{6} \} = 6
\end{equation}
\begin{equation}
2d6kl=\{ \vcdice{3}, \vcdice{6} \} = 3
\end{equation}
\begin{equation}
2d6dh=\{ \vcdice{3}, \vcdice{6} \} = 3
\end{equation}
\begin{equation}
2d6dl=\{ \vcdice{3}, \vcdice{6} \} = 6
\end{equation}

Negative values of z produce an error as the requested behaviour is ambiguous.




When z is the same or larger than the number of dice as it is in Equation \ref{eq:sdk3}, there is no effect.

\begin{equation}
\label{eq:sdk3}
2d6kh3=\{ \vcdice{2}, \vcdice{3}\} = \{2,3\} = 5
\end{equation}

A less common use case is to select the middle dice. Rather than define a new \textit{m} case which would need unintuitive rules for handling both odd and even numbers of dice (or mixed rules like "upper-middle" or "lower-middle"), we propose that this rare case be handled by dropping both sides instead. An example of choosing the middle value with this method is shown in Equation \ref{eq:sdk4}

\begin{equation}
3d6dldh=\{ \vcdice{3}, \vcdice{6}, \vcdice{4} \} = 4
\label{eq:sdk4}
\end{equation}

\paragraph{Conditions \& Filters}
\label{filters}

In many games, certain logic is only required if a particular value is rolled. For example, the board game One Deck Dungeon \cite{ODD} contains a scenario that requires the player to discard any 2s rolled in their roll. Conditional statements allow a high range of flexibility for cases such as these.

When applied to a vector of rolls, a conditional can act as a filter to remove rolls that do not satisfy the criteria. 

Conditionals in GNOLL operations take the same form as one might find in typical programming languages (i.e. ==, !=, <, >, <=, >=).

Filters use conditionals to remove dice rolls that do not match the conditional check. Equation \ref{eq:filter} shows four dice being rolled and then two of them discarded for not matching the filter criteria.

\begin{equation}
4d6f<3 = filter(\{ \vcdice{4},  \vcdice{1}, \vcdice{2}, \vcdice{5}\}, "<3") =  \{ \vcdice{2}, \vcdice{1} \}
\label{eq:filter}
\end{equation}


\paragraph{Rerolling}
\label{rerolling}

Rerolling dice is often used to avoid unfavorable conditions in TTRPGs, for example, it can be used to ensure that players do not create characters with statistics that are significantly lower than their peers. In Dungeons \& Dragons \cite{PHB}, Halfling player characters have the natural ability to reroll dice that have a result of 1. 

Arguably, some of these rolls could be written in a more computationally efficient manner (e.g. instead of rerolling 1s on a d20, one could roll a single dice with values from 2 to 20 instead), but TTRPG game designers design primarily for physical games (at least, at the time of writing) and it makes sense to keep aligned to the paradigms of the hobby for greater accessibility.

Rerolling is configurable in GNOLL. a single \lq r\rq\  will reroll a single time (as shown in Equation \ref{eq:ronce}). This can be optionally followed with another \lq r\rq\ to continue rerolling until the triggering condition is no longer fulfilled (As in Equation \ref{eq:rall}. Internally, there is a hardcoded limit so that execution does not continue infinitely. 


\begin{equation}
1d6r<2 = reroll(\{ \vcdice{4} \}, "once", "if <2") =  \{ \vcdice{2} \rightarrow{} \vcdice{1} \}
\label{eq:ronce}
\end{equation}
\begin{equation}
1d6rr<2 = reroll(\{ \vcdice{4} \}, "\infty", "if <2") =  \{ \vcdice{2} \rightarrow{} \vcdice{1} \rightarrow{} \vcdice{4} \}
\label{eq:rall}
\end{equation}

\paragraph{Explosions}

In some systems like Feng Shui \cite{fengshui2}, dice can \lq explode\rq{}, which is to say that if a die rolls its maximum value (e.g. a 6 for a d6), it rolls again and accumulates the new roll. The most popular syntax for this is an exclamation point as shown in Equation \ref{eq:exp}. Occasionally the lowercase \lq e \rq is used. 

\begin{equation}
1d6! = explode(\{ \vcdice{6} \}, N) =  \{ \vcdice{6} \rightarrow{} \vcdice{6} \rightarrow{} \vcdice{4} \} = 16
\label{eq:exp}
\end{equation}

There is a hardcoded limit on exploding in GNOLL, but it's unlikely to be met. Two roll types reduce impact of rolling multiple times when playing with physical dice --- only allowing a single explosion (!o --- o for \lq once\rq\ ) or using \lq penetrating\rq\ exploding dice (!p) in which the value of the accumulation decreases by one each explosion. When the potential addition is 0, rolling can cease. Worked examples of each can be seen at Equations \ref{eq:expone} and \ref{eq:rall}.

\begin{equation}
1d6!o = explode(\{ \vcdice{6} \}, 1) =  \{ \vcdice{6} \rightarrow{} \vcdice{6} \} = 12
\label{eq:expone}
\end{equation}

\begin{equation}
\begin{aligned}
1d6!p = explode(\{ \vcdice{6} \}, p) \\
= \{ \vcdice{6} \rightarrow{} \vcdice{6} \rightarrow{} \vcdice{6} \rightarrow{} \vcdice{2} \} \\ = (6+(6-0)+(6-1)+2) = 19
\end{aligned}
\label{eq:exppen}
\end{equation}


\paragraph{Success Counting}
Shadowrun\cite{shadowrun} and some other systems forgo complex single-dice mechanics for a simpler method of rolling a larger number of dice and counting how many score above a given threshold.

GNOLL provides a count token \lq c\rq\ which can be applied after a filter to produce this effect as shown in \ref{eq:count}. In the following Equation \ref{eq:uc} We also introduce a new built-in 'u' which is a filter that removes any duplicate values, allowing a user to count unique rolled results.

\begin{equation}
4d6f>2c = count( filter(\{ \vcdice{4},  \vcdice{1}, \vcdice{2}, \vcdice{5}\}, "!=2") ) =  3 \}
\label{eq:count}
\end{equation}
\begin{equation}
4d6uc = count(unique( \{ \vcdice{4},  \vcdice{1}, \vcdice{2}, \vcdice{5}\}, u) ) =  4 \}
\label{eq:uc}
\end{equation}

\subsubsection{Mathematical Operations}

Overall, dice math in GNOLL is similar to usual infix mathematical notation. Before operations are complete, sets of dice are collapsed into their smallest form. Where multiple values continue to exist, operations are applied element-wise. Where dimensions of the vectors do not match, default values are used in their place to best preserve a value (e.g. Equation \ref{element-wise}).

\begin{equation}
d6 + d6=\{ \vcdice{3}\}+\{ \vcdice{6} \} = 9
\end{equation}
\begin{equation}
\label{element-wise}
(d6;d6) - 3 = \{\{ \vcdice{3}\},\{ \vcdice{6} \}\}-3 = \{0,6\}
\end{equation}
\begin{equation}
(d6;d6) \times{} (d6;d6) = \{\{ \vcdice{3}\},\{ \vcdice{6} \}\} \times{} \{\{ \vcdice{2}\},\{ \vcdice{4} \}\} = \{ 6, 24 \} 
\end{equation}

An alternative to this could be to use vector and matrix math, but to be simpler, this has been reserved for future work should there be appropriate requests for such a feature.

TTRPGs and board games predominately operate with integers only. Therefore, it may be useful to support two types of division supporting different types of rounding modes. GNOLL implements these as backslash (Equation \ref{eq:rdown} and forwardslash (Equation \ref{eq:rup}).

\begin{equation}
3 / 2 =  1
\label{eq:rdown}
\end{equation}
\begin{equation}
3 \backslash 2 = 2
\label{eq:rup}
\end{equation}

\subsection{Programmatic Operations}

\paragraph{Macros}
\label{macros}

If you have a complex or wordy symbolic dice, or some very atypical logic - dice rolls can be lengthy to write --- especially if you need to reuse those portions multiple times. Macros resolve this issue by storing the dice configuration in a lookup table, referred to by name. GNOLL maintains a list of  useful macros for common symbolic dice (e.g. Chess Dice, Poker Dice). Users can submit their macros for consideration for inclusion in the repository. 

An example of setting and accessing a macro are given below in Equations  and \ref{eq:macro_access}. Macro statements that save expressions (e.g. Equation \ref{eq:macro_set}) and access them (e.g Equation \ref{eq:macro_set}) can be chained using a semicolon separator.

\begin{equation}
\label{eq:macro_set}
\#SUITS = d\lbrace CLUBS, HEARTS, DIAMONDS, SPADES \rbrace
\end{equation}

\begin{equation}
\label{eq:macro_access}
@SUITS
\end{equation}
\section{Results}
\label{results}

\subsection{Performance Recordings}
Unfortunately, as many popular dice notation software solutions are either proprietary or web-based, it is difficult or impossible to fairly compare their performance to GNOLL.

However, where implementations of standalone dice rolling libraries existed, we were able to profile their performance to benchmark against GNOLL. Where compared software produced an overly verbose output (i.e. printing the 100 die results of a 100d100), the code was modified to remove the output in order to do a fair comparison.

\subsubsection{Performance comparison versus TROLL and C++ library Dice Parser}

We profiled GNOLL's native interface (at version \hyperlink{https://github.com/ianfhunter/GNOLL/releases/tag/v2.2.3}{v2.2.3}) against the C++ library \hyperlink{https://github.com/EBailey67/DiceParser}{Dice Parser by Eric D. Bailey}.

We also compared performance against \hyperlink{http://hjemmesider.diku.dk/~torbenm/Troll/}{TROLL}. It is written using MoscowML, an implementation of the StandardML(SML) programming language. 

The results are displayed in Figure \ref{fig:c_perf} where it can be seen GNOLL consistently out-performs the others at each step.

\pgfplotstableread[col sep=comma]{CDATA.csv}\datatable

\begin{figure}[h]
\begin{tikzpicture}
\begin{axis}[
    xlabel=Dice Roll (NdN),
    ylabel=Time (s),
    flexible xticklabels from table={CDATA.csv}{category}{col sep=comma},
    xticklabel style={text height=1.5ex}, 
    xtick=data,
    log ticks with fixed point,
    ytick={1,10,100,1000,10000,100000},
    ymode=log,
    legend pos = north west
]
\addplot table[x expr=\coordindex,y=gnoll_perf]{\datatable};
\addlegendentry{GNOLL} 

\addplot table[x expr=\coordindex,y=troll_perf]{\datatable};
\addlegendentry{TROLL} 

\addplot table[x expr=\coordindex,y=ok_perf]{\datatable};
\addlegendentry{Dice Parser} 

\end{axis}
\end{tikzpicture}
\caption{Performance Comparison of C/C++/SML Implementations}
\Description{A line chart showing the performance of GNOLL, TROLL, and Dice Parser. All lines move upwards as the operations get more complex, but GNOLL performs significantly better than the others. TROLL has a much sharper incline than the other two items.}
\label{fig:c_perf}
\end{figure}
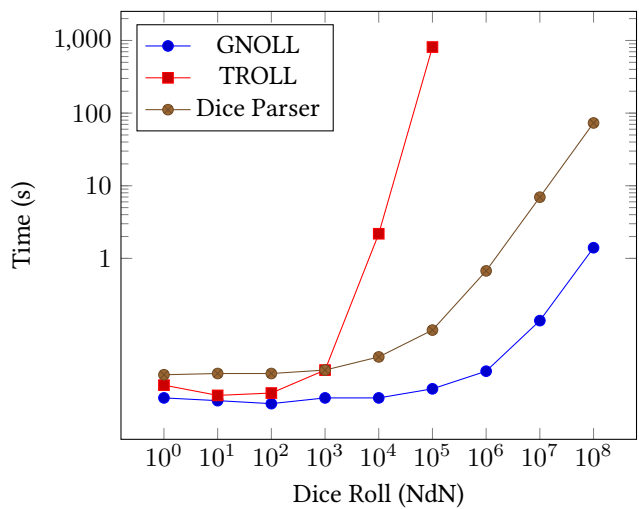

\begin{table}[h]
  \caption{Library Support Matrix}
  \label{table:c}
  \begin{tabular}{lcccc}
    \toprule
    Feature & GNOLL & TROLL & Dice Parser \\
    \midrule
     system  & CFG & CFG & CFG  \\
     basic xDy & \checkmark & \checkmark & \checkmark \\
     fate dice & \checkmark &  \text{\sffamily X} & \checkmark \\
     math & \checkmark & \checkmark & \checkmark \\
     drop/keep & \checkmark & \checkmark & \checkmark\\
     explosions & \checkmark & \checkmark & \checkmark \\
     rerolling & \checkmark & \checkmark & \checkmark \\
     filtering & \checkmark & \checkmark & \text{\sffamily X}\\
     distributions & \text{\sffamily X} & \checkmark & \text{\sffamily X} \\
  \bottomrule
\end{tabular}
\end{table}

\pgfplotstableread[col sep=comma]{testdata.csv}\datatable

\subsubsection{Performance comparision against popular Python parsers}

PyPi \cite{pypi} is a popular repository of Python packages that many use to distribute their software. We measured several dice parsing libraries from PyPi's catalogue against GNOLL's Python3 interface.

The exact libaries installed via installation tool `pip' and benchmarked are: 
\begin{itemize}
    \item \hyperlink{https://pypi.org/project/dice/}{dice},  
    \item \hyperlink{https://pypi.org/project/rpg-dice/}{RPG Dice}
    \item \hyperlink{https://pypi.org/project/python-dice/}{Python Dice}, 
\end{itemize}

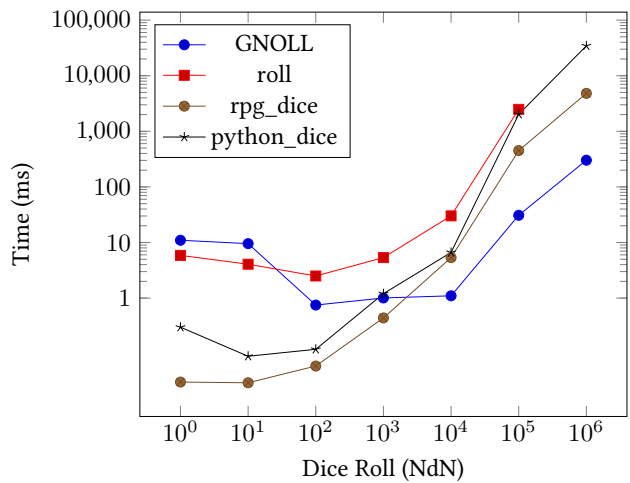
\begin{figure}[h]
\resizebox{\columnwidth}{!}{
    \begin{tikzpicture}
    \begin{axis}[
        xlabel=Dice Roll (NdN),
        ylabel=Time (ms),
        flexible xticklabels from table={testdata.csv}{category}{col sep=comma},
        xticklabel style={text height=1.5ex}, 
        xtick=data,
        log ticks with fixed point,
        ytick={1, 10,100,1000,10000,100000},
        ymode=log,
        legend pos = north west
    ]
    \addplot table[x expr=\coordindex,y=gnoll_perf]{\datatable};
    \addlegendentry{GNOLL} 
    
    \addplot table[x expr=\coordindex,y=roll_perf]{\datatable};
    \addlegendentry{roll} 
    
    \addplot table[x expr=\coordindex,y=rpg_perf]{\datatable};
    \addlegendentry{rpg\_dice} 
    
    \addplot table[x expr=\coordindex,y=pil_perf]{\datatable};
    \addlegendentry{python\_dice} 
    
    \end{axis}
    \end{tikzpicture}
}

\caption{Performance Comparison of Python Dice Notation Packages}
\Description{A line chart comparing the performance of GNOLL against several python packages. GNOLL starts as the slowest of the four by a small margin, but as the complexity increases, completes the fastest.}
\label{fig:py_perf}
\end{figure}

The results are displayed in Figure \ref{fig:py_perf}. GNOLL appears to have good long-running efficiency, but for short commands falls short (Though, the overall time is much shorter and would likely not be noticeable by a user).

\begin{table}[h]
  \caption{Python Library Support Matrix}
  \label{table:py}
  \begin{tabular}{lcccc}
    \toprule
    Feature & GNOLL & roll & rpg\_dice & python\_dice \\
    \midrule
     system  & CFG & CFG & RegEx & CFG  \\
     basic xDy & \checkmark & \checkmark & \checkmark & \checkmark  \\ 
     fate dice & \checkmark & \checkmark & \text{\sffamily X} & \checkmark   \\ 
     math & \checkmark & \checkmark & \checkmark & \checkmark  \\ 
     drop/keep & \checkmark & \checkmark & \text{\sffamily X} & \checkmark  \\ 
     explosions & \checkmark & \checkmark & \text{\sffamily X} & \text{\sffamily X}  \\ 
     rerolling & \checkmark & \checkmark & \text{\sffamily X} & \text{\sffamily X}  \\ 
     filtering & \checkmark & \text{\sffamily X} & \text{\sffamily X} & \text{\sffamily X}  \\ 
     distributions & \text{\sffamily X} & \text{\sffamily X} & \text{\sffamily X} & \checkmark  \\ 
  \bottomrule
\end{tabular}
\end{table}

\subsection{Examples of Supported TTRPGs}
There are an innumerable amount of TTRPGs, but many of them share similar rules and only differ slightly or have setting changes (e.g. \lq Pathfinder \rq\ is set in a Fantasy setting, whereas \lq Starfinder\rq\ is set in a Science Fiction setting). Below is a sample of systems supported by GNOLL's current feature set:

\begin{itemize}
    \item D20 Systems --- Dungeons \& Dragons, Pathfinder, Dungeon Crawl Classics
    \item 2D20 Systems --- Star Trek Adventures, Conan: Adventures in an Age Undreamed of
    \item Powered By The Apocolypse --- Dungeon World, Monster of The Week
    \item Fate Systems --- Fate Core, Fate Accelerated
    \item TinyD6 --- Tiny Dungeon, Knights of Underbed
    \item Star Wars RPG \footnote{There are multiple Star Wars RPGs, The one tested was developed by Fantasy Flight Games}
\end{itemize}

\section{Discussion}

\subsection{Impact of Results}

As mentioned in Section \ref{results}, it was not possible to benchmark against some popular TTRPG dice rolling software as they were inaccessible or prone to external factors (e.g. network latency). However, from our sampling of various other dice notation systems, it can be seen that GNOLL performs well in comparison --- especially at higher complexities.

One anomaly in our result gathering is in our Python results, a roll of 10000d10000 is recorded to be faster than a roll of 1d1. As both times are still quite small, we assume that this unexpected behavior may be due to the binding system itself as we do not observe this behavior in the C executable.

The feature matrices in Tables \ref{table:c} and \ref{table:py} show a rough negative correlation between feature support and speed (TROLL, roll being the most feature-full, but also the slowest implementations). GNOLL manages to subvert this correlation.

The main absence in these tables for GNOLL is reporting a dice roll's statistical distribution. This is penned for future development, but is expected to be a secondary feature only used by developers or advanced users.

Overall, GNOLL appears to be a comparatively efficient library for parsing dice notation, as well as being feature-rich. 

\subsection{Limitations}

\subsubsection{Pre/Post Processing}

As mentioned in Section \ref{scope}, GNOLL has defined a narrow scope for development and so has some self-enforced limitations, mostly around post/pre-processing. Some systems like D\&DBeyond connect character sheets to their dice notation so that when a value is changed on the sheet (e.g. the character gains a level), the dice roll is updated, but there is no change of workflow to the user. Figure \ref{cleo_dex_save} shows a Roll20 dice roll \cite{roll20_dice} --- using this notation, the player controlling the character Cleo would not have to change the syntax should their character sheet be updated to indicate equipping a magic item or some other bonus conferral to the Dex-mod value. 

Due to GNOLL's scope limitation, this must be done in a preprocessing step by the host software. 


\begin{figure}[H]
    \centering{}
    \begin{mdframed}
\begin{lstlisting}[columns=fullflexible]
           /roll 1d20+@{Cleo|DEX-mod}
\end{lstlisting}
    \end{mdframed}
    \caption{Roll20 notation for using a character sheet property}
    \label{cleo_dex_save}
\end{figure}

Similarly, contextual information like highlighting "Critical Failures" or "Critical Successes" (A minimum or maximum roll of a d20 dice in Dungeons \& Dragons) must be done in post-processing.

\begin{figure}[H]
  \includegraphics{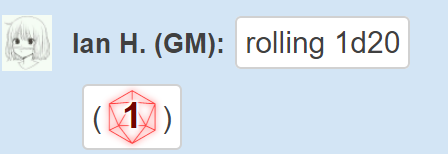}
  \caption{Superficial display of a critical failure in Roll20.}
  \Description{The number 1 is highlighted in a red stylized hexagon to distinguish itself from a normal roll}
  \label{fig:critfail}
\end{figure}





\subsubsection{Programmatic Freedom}
GNOLL is not a programming language and currently lacks some of those common constructs that are present in machine-friendly DSLs. Aspects of some are present - Conditionals and rerolling allow a certain degree of similarity to While and If statements, but writing complex atypical logic within GNOLL's syntax is sure to encounter difficulties. However, we do not foresee many legitimate cases of this being required within the TTRPG domain.

\subsubsection{Degrees of Success}

Certain TTRPG games like Fate\cite{FateCore} have degrees of success depending on the roll result. For example, one might achieve a \lq Partial Success\rq\  or a \lq Extraordinary Success\rq\  depending on how significantly their dice exceeded the threshold. 

GNOLL provides functionality to count the number of dice in a result. When combined with filters, this can act as a success/failure count for a player. However, GNOLL is limited in that you can currently only retrieve a single count from a roll at a time. This is a side-effect of the limited complexity of GNOLL and it is recommended for users desiring this behavior to retrieve the full set of results from GNOLL and post-process them to apply logic to classify their result.

\subsection{Future Work}
\label{future_work}

GNOLL is an ongoing project and it is expected to have a continuous backlog of development work for the foreseeable future. There are some items that we see on the immediate horizon:

\subsubsection{Language Support}

At present, GNOLL maintains a C, C++, Perl and Python interface. As mentioned in Section \ref{code_architecture}, SWIG provides many bindings from C to other languages, but while it might be possible for users to create their bindings it would be more convenient for them to use bindings from a GNOLL release that has been tested and verified. At present, a SWIG interface file is provided, but bindings for other languages SWIG supports are not created yet.

We are particularly interested in expanding GNOLL support to languages known for web development, though we welcome any other binding contributions.

\subsubsection{Additional Features}

GNOLL initially focussed on TTRPG dice, though there are many dice-based board games and digital games that can be considered as inputs for its feature development.

\paragraph{Substitution}
The board game One Deck Dungeon \cite{ODD} follows an adventurer as they delve through various floors of a dungeon. On some floors, the player may be told to force certain rolls to be changed to a set value. A prospective syntax follows in Equation \ref{eq:subs}
\begin{equation}
\label{eq:subs}
4d6set<2to3 = \{ \vcdice{4},  \vcdice{1} \rightarrow{} \vcdice{3}, \vcdice{2}, \vcdice{5}\}
\end{equation}

\paragraph{Expanded Macros \& Shorthand}
dF.1 is sometimes used for a fate dice with only one minus and one plus side (all others are blank). In this notation, the usual dF is known as dF.2. This could be extended to be complete with dF.3 where there are no blank sides, but three minus symbols and three plus symbols.

d10x is used in the Cyborg Commando TTRPG \cite{Commando} and is shorthand for d10 * d10.

\lq Block Dice\rq\ of the form $ Xdb $ are used by Blood Bowl games and expand to $Xd6kh1$ --- i.e. roll X d6 and choose the highest value.

\paragraph{Diminishing Explosion}
A variant of penetrating explosion called diminishing explosion reduces the value of the dice sides rather than the value. For games with a standard 7-dice set (d\%\footnote{see Section \ref{special_dice}}, d20, d12, d10, d8, d6, d4) players roll the next smallest dice (e.g. after rolling a maximum 20 on a d20, the player would roll a d12). In this case, we would expect users to specify the dice array as an input as the implementation is dependent on what set of dice a system uses.  A prospective syntax follows in Equation \ref{cascading}.

\begin{equation}
\label{cascading}
\begin{aligned}
(1d6;1d4)!p = explode(\{ \vcdice{6} \}, p) \\
= \{ \vcdice{6} \rightarrow{} \vcdice{4} \}  = 10
\end{aligned}
\end{equation}

\subsection{Future Application}
\label{future_applications}
We hope that, like TROLL and its predecessor, GNOLL will be used as a platform for future research. We would like to expand GNOLL's supplemental scripts to offer support for this type of work (e.g. displaying probability distributions, optimization of queries, etc).

As the support for more languages and GNOLL's reputation increases, we hope to have more users and contributors to the GNOLL project.

One of the longer term goals of GNOLL is to encourage an agreed non-trivial dice notation standard across the TTRPG and gaming domains. 

\subsection{Conclusion}

There is a gap in the open-source software world for a dice notation parser that is both easy to use and easy to integrate. In this paper, we introduced GNOLL --- an efficient dice rolling library based on a dice notation DSL and built upon Unix tools YACC and LEX. 

There is a lack of standardization between various dice notation implementations in popular TTRPG software and GNOLL's syntax is based on those conventions which are established or by independent analyses of appropriateness for use.

GNOLL is available on GitHub for download and contribution. \href{https://github.com/ianfhunter/GNOLL}{https://github.com/ianfhunter/GNOLL}

\begin{acks}
Thanks to my wife, for her patience --- and to my dog, for her impatience. \\
To the various TTRPG publishers and online communities for enabling people all over the world to play make-believe together.
\end{acks}

\bibliographystyle{ACM-Reference-Format}
\bibliography{main}

\appendix
\pagebreak{}
\section{Appendix}

\subsection{Grammar Overview}
\label{ax:grammar}

Grammar diagrams generated using Tabatkin's Railroad Diagram Generator (\url{https://tabatkins.github.io/railroad-diagrams/}).

These diagrams are correct as of time of writing.

\begin{figure*}[h]
  \includegraphics[width=\textwidth]{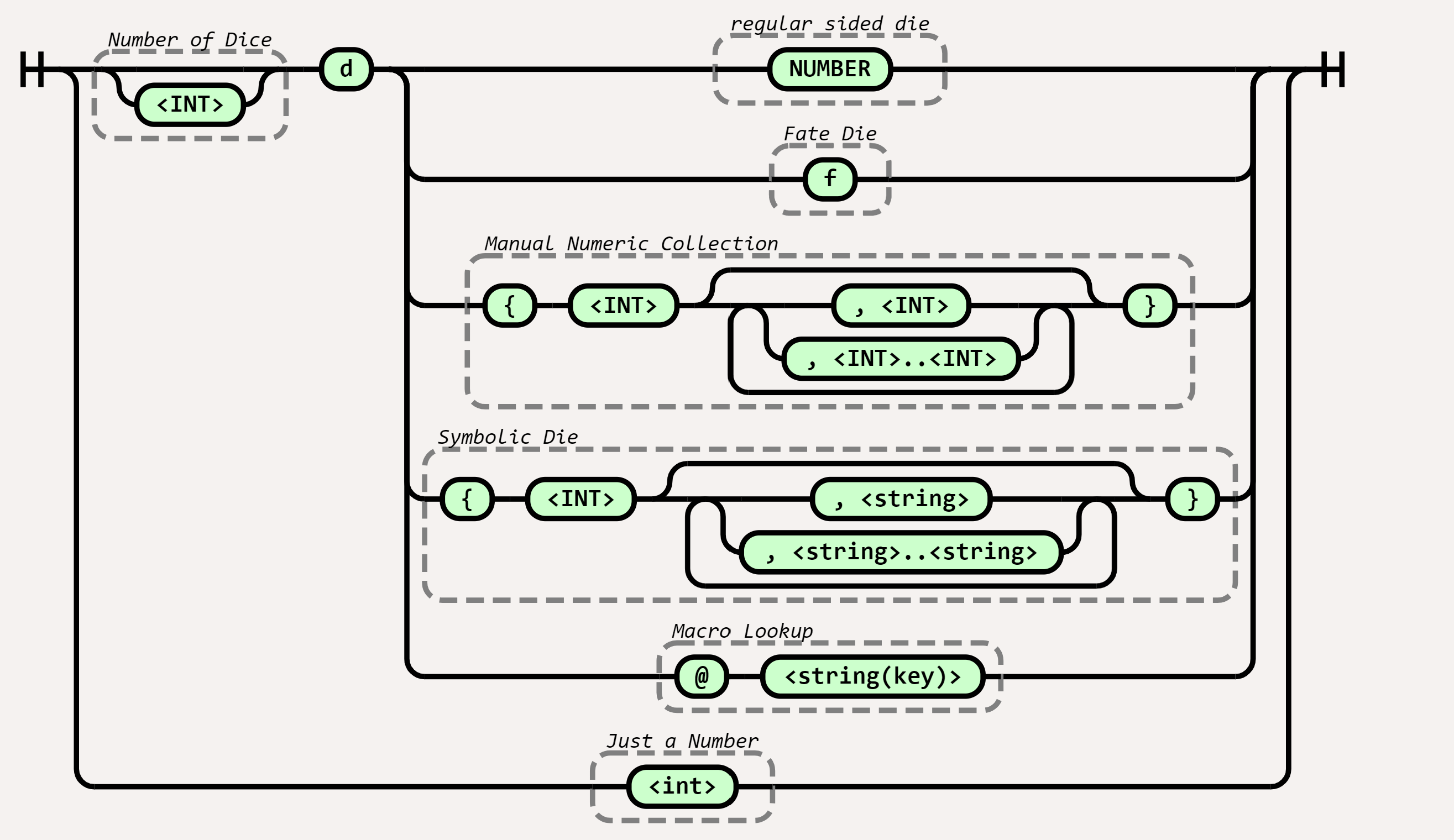}
  \caption{Railroad Diagram of Dice Terminals}
  \Description{}
  \label{fig:grammar:dice}
\end{figure*}

\begin{figure*}[h]
  \includegraphics[width=0.5\textwidth]{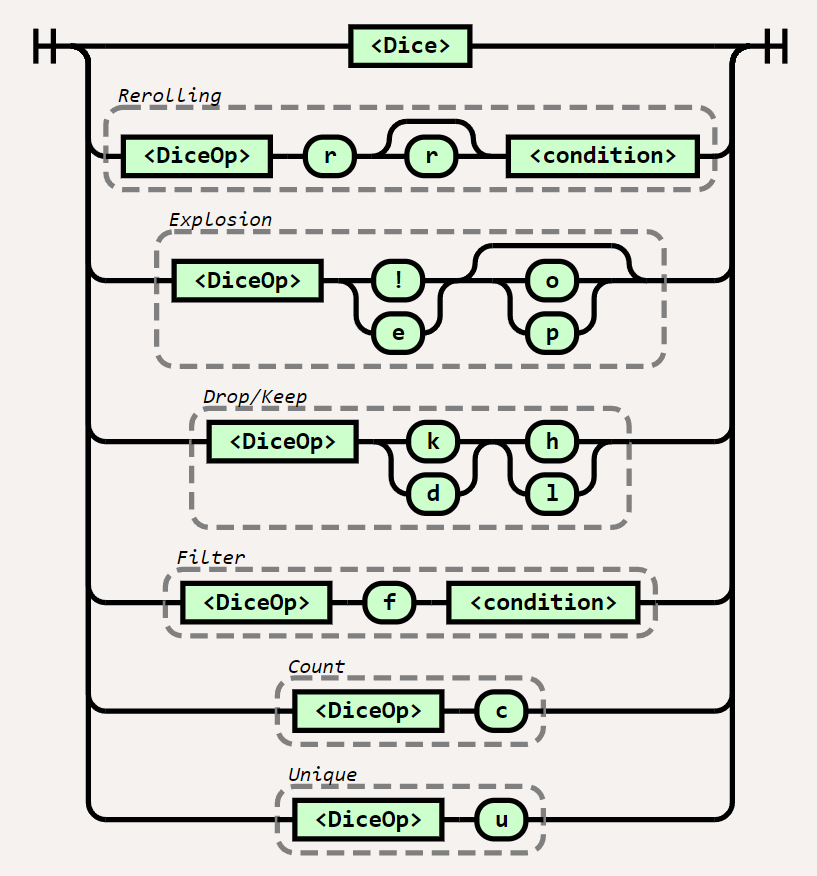}
  \caption{Railroad Diagram of dice operations}
  \Description{}
  \label{fig:grammar:dice_op}
\end{figure*}

\begin{figure*}[h]
  \includegraphics[width=0.5\textwidth]{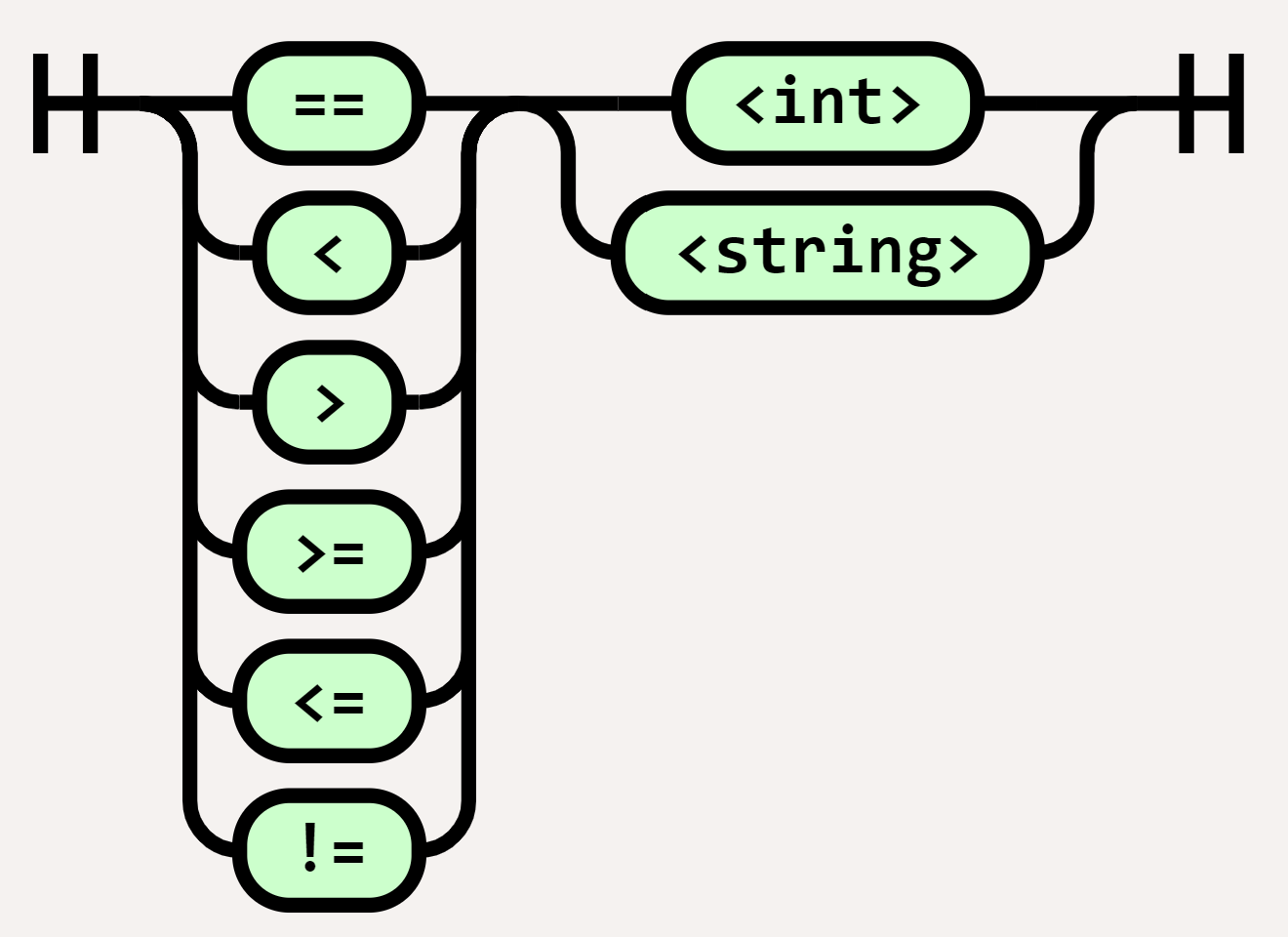}
  \caption{Railroad Diagram of conditions used in filters}
  \Description{}
  \label{fig:grammar:dice_cond}
\end{figure*}

\begin{figure*}[h]
  \includegraphics[width=0.5\textwidth]{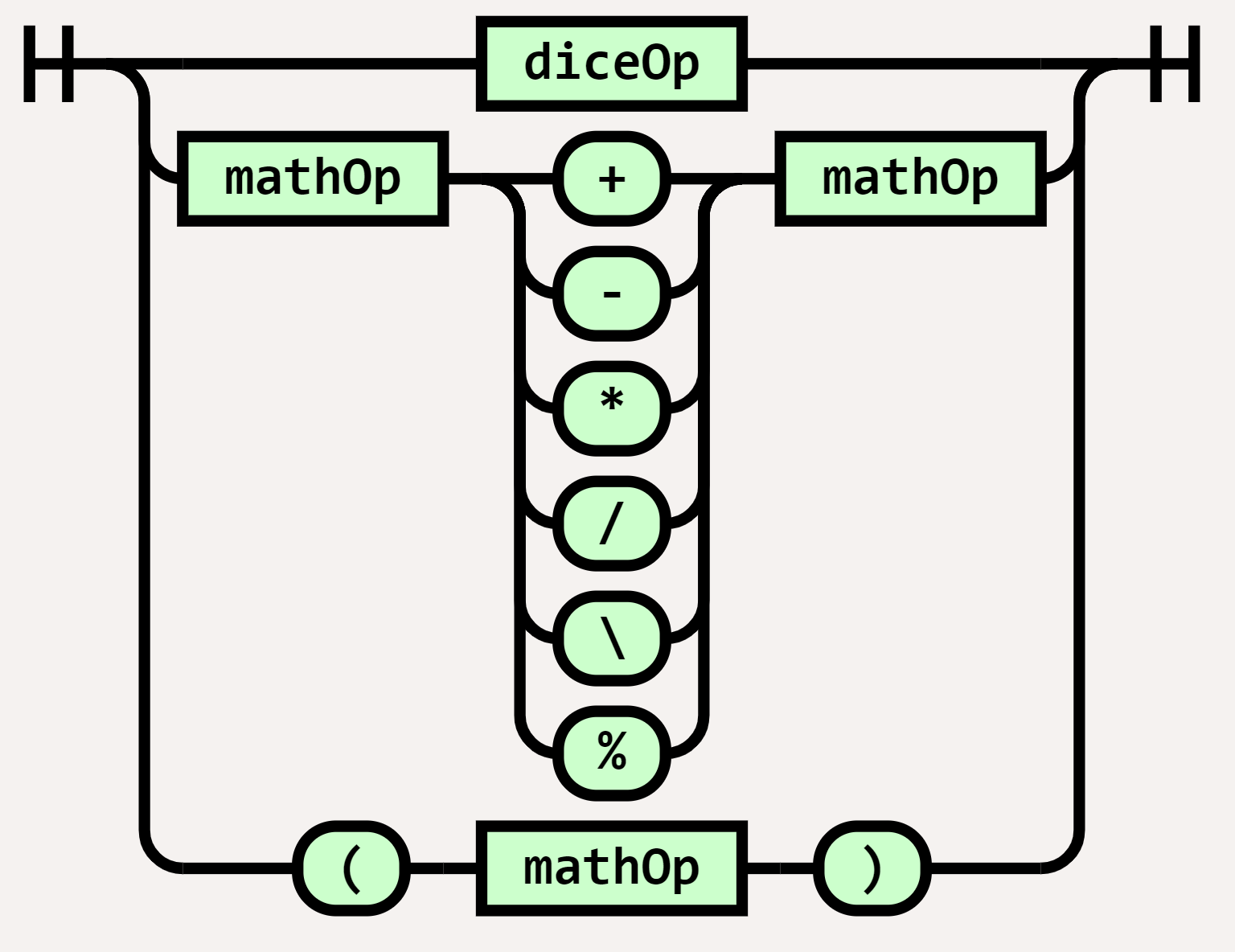}
  \caption{Railroad Diagram of math operations}
  \Description{}
  \label{fig:grammar:math}
\end{figure*}

\begin{figure*}[h]
  \includegraphics[width=\textwidth]{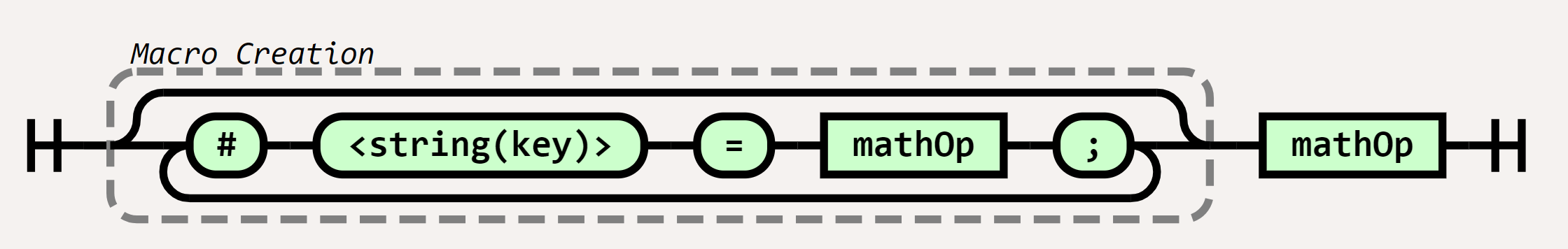}
  \caption{Railroad Diagram of programmatic operations}
  \Description{}
  \label{fig:grammar:prog}
\end{figure*}

\end{document}